# *Carica papaya*'s Leaf Extract Inhibits SARS-CoV-2 Main Proteases but not Human TMPRSS2:

# An *In-vitro* and *In-silico* Study

Maywan Hariono[1,3*], Irwan Hidayat[2], Ipang Djunarko[3], Jeffry Julianus[3], Fadi G. Saqallah[4], Muhammad Hidhir Khawory[5], Nurul Hanim Salin[5], Habibah Abdul Wahab[4]

[1] *Pharmaceutical Chemistry Department, School of Pharmaceutical Sciences, Universiti Sains Malaysia, 11800 Minden, Pulau Pinang, Malaysia*

[2] *PT Industri Jamu dan Farmasi Sido Muncul Tbk, Soekarno Hatta Street Km 28, Klepu, Semarang 550552, Jawa Tengah, Indonesia*

[3] *Faculty of Pharmacy, Sanata Dharma University, Maguwoharjo, Depok, Sleman 55282, Yogyakarta, Indonesia*

[4] *Pharmaceutical Technology Department, School of Pharmaceutical Sciences and USM-RIKEN Centre for Ageing Science (URICAS), Universiti Sains Malaysia, 11800 Minden, Pulau Pinang, Malaysia*

[5] *Malaysian Institute of Pharmaceuticals and Nutraceuticals, National Institute of Biotechnology Malaysia, Halaman Bukit Gambir, 11900 Bayan Lepas, Pulau Pinang, Malaysia*

**\* Correspondence:** *Maywan Hariono, Ph.D.; Pharmaceutical Chemistry Departement, Universiti Sains Malaysia, 11800 Minden, Pulau Pinang, Malaysia; Phone: +604-6532211; Fax: + 604-6570017; Email:* maywanhariono@usm.my

## ABSTRACT

*Carica papaya* (CP) leaf is long known for its traditional pharmacological effects against dengue virus and malaria. Therefore, CP could also be a potential solution for the treatment of other infectious diseases, such as coronavirus. In this study, we evaluate the potential effect of the ethanolic CP leaf extract in inhibiting the enzymatic activity of three protein targets in SARS-CoV-2's life cycle, which include the 3-chymotrypsin-like protease (3CLpro), papain-like protease (PLpro) and the human transmembrane protein serine 2 (TMPRSS2). Results demonstrate that CP's leaf extract inhibits 3CLpro and PLpro significantly with $IC_{50}$'s of 0.02 μg/mL and 0.06 μg/mL, respectively, but it is inactive towards TMPRSS2. Phenol, 2-methyl-5-(1,2,2-trimethylcyclopentyl)-(*S*)- (**17a**) and β-mannofuranoside, farnesyl- (**21a**) were identified in the extract using GC-MS. These two compounds demonstrated a stronger binding affinity towards the main proteases than TMPRSS2 during the docking simulation, which agrees with the *in-vitro* study. Further pharmacophore mapping suggests that **17a** has a fit score higher than **21a** to the SARS-CoV-2 3CLpro pharmacophore model concluding that CP leaf extract has the potential to be developed as a herbal SARS-CoV-2 antiviral agent.

# 1. INTRODUCITON

Papaya (*Carica papaya* L.) originates from Southern Mexico or Northern South America and is widely planted in the whole tropical region to be utilized for its fruit [1]. The fruit is highly consumed by people due to its non-seasonal production. Along with their knowledge and understanding, people not only consume papaya as a dessert but also for its health-promoting effects [2]. The main components in papaya are papain and chymopapain, mainly used as textile materials [3], whereas the other phytochemical components, such as alkaloids, flavonoids, and phenols, were reported for their activity towards malaria, dengue, and diabetes mellitus [4].

An unpublished study from our laboratory found that the ethanol extract of CP leaf was able to inhibit SARS-CoV-2 3CLpro enzymatic activity by 64% at 1000 µg/mL. The 3CLpro is a proteolytic enzyme playing a crucial role in the SARS-CoV-2 life cycle as it hydrolyses the polyprotein in the eleven conserved sites of SARS-CoV-2 [5]. 3CLpro is a non-structural protein 5 (NSP5) having a catalytic dyad composed of cysteine and histidine residues that cleave the substrate at the Gln–(Ser/Ala/Gly) peptide bond [6]. The hydrolysed product would be assembled with the structural proteins to generate a sub-genome through discontinuous transcription from the 5' end of the anti-sense viral RNA [7]. After successful genome replication and translation, the NSPs, structural proteins, and some accessory proteins assemble along with the positive-sense viral RNA genome to form a new virion [8].

The next main protease in coronavirus is the papain-like protease (PLpro), which recognizes and hydrolyses a cellular ubiquitin (Ub) and the UBL ISG15 proteins, an interferon induced by gen 15 [9]. Both proteins act to recognize the LXGG peptide at the C-terminal motif. Ub and ISG15 are important in cellular modification, where they both bind covalently to the targeted protein via isopeptidic bond formation [10]. This isopeptidic bond could be hydrolysed by the isopeptidase enzymatic activity from DUb and the deISG15 to cleave Ub and ISG15 from the host cell protein [11]. In the case of SARS coronavirus, PLpro has been studied to interact as a cellular response antagonist, which depends on Ub upon viral infection. Although the mechanism on how PLpro mediates this cellular antagonism has not been fully understood, limited experimental evidence suggests that the catalytic activity is essential in the antagonism. Therefore, DUb and deISG15 are the proposed mechanism [12].

Much earlier, as the SARS coronavirus invades the host cell, a protein, namely transmembrane protein serine 2 (TMPRSS2), plays a crucial role in facilitating the virus attachment to the host cell by binding to angiotensin-converting enzyme-2 (ACE-2) [13, 14]. This attachment is mediated by the coronavirus's structural spike (S) protein, underlining that TMPRSS2 is another strategic protein target. Also, this enzyme is included in the proteases family, expressed from the luminal part of normal prostatic epithelium, and developed by a malignant prostate tissue [15]. Furthermore, TMPRSS2 in the avian eritoblastotic virus shows homology with the E26 oncogene (ERG) that is present in 50% of prostate tumor cases in the Caucasian male and is often early detected in prostate carcinogenesis [16]. TMPRSS2 and ERG genes are tandemly located at the 21q22 chromosome, and the expression of TMPRSS2 is regulated by androgens [17, 18].

Those three proteins are available as 3D structures in complex with small inhibitors, which would help in studying the molecular mechanism of how small molecules inhibit their corresponding enzymatic activity. **Fig. 1a** visualizes baicalein, a flavonoid, bound to the SARS-CoV-2 3CLpro [19], whereas **Fig. 1b** shows the binding of a peptidomimetic compound, namely VIR251, to the SARS-CoV-2 PLpro [20], and **Fig. 1c** demonstrates the binding of nafamostat, a carbaimidebenzoic acid, to the human TMPRSS2's active site [21]. Any compound with this kind of binding mode would potentially serve as an inhibitor to the respective enzyme.

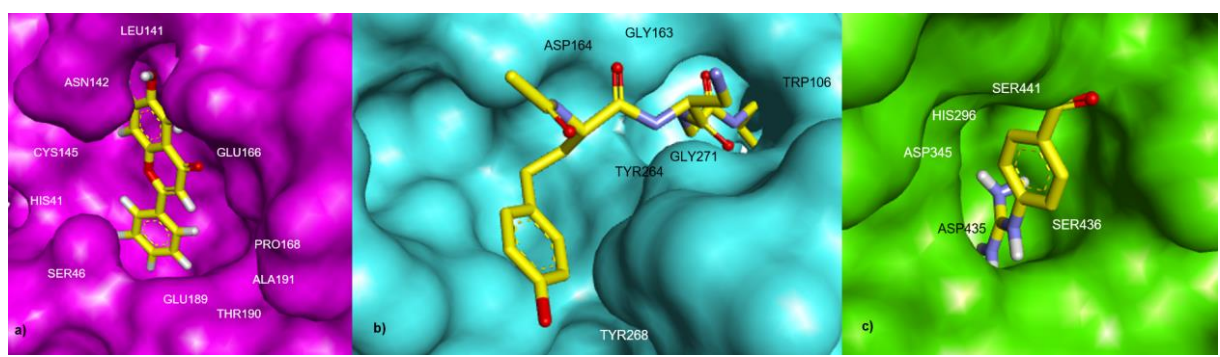

**Fig. 1** The 3D complex structures of **a)** SARS-CoV-2 3CLpro – baicalein, **b)** SARS-CoV-2 PLpro – VIR251, and **c)** human TMPRSS2 – nafamostat, with their surrounding amino acid residues important for their binding and activity.

Next to the 3CLpro, our unpublished study also showed that 1000 µg/mL CP's leaf ethanolic extract was able to inhibit PLpro and TMPRSS2 by 57% and 21%, respectively. In this study, we are motivated to explore further the lowest concentrations needed to inhibit at least 50% of the corresponding enzyme activity, also known as the $IC_{50}$, and the extract's safety via a cytotoxicity study employing MTT assay in Vero cells. The CP extract was then characterized for its chemical composition using gas chromatography-mass spectrometry (GC-MS) and liquid chromatography-mass spectrometry (LC-MS). The identified major compounds were then studied for their enzyme binding affinities using computational molecular docking and structure-based pharmacophore mapping.

## 2. RESULTS AND DISCUSSION

*In-vitro*, the lowest concentration of CP leaf extract (62.5 µg/mL) showed up to 34% inhibition towards the 3CLpro. Herein, a dose-dependent relationship was noticed where the inhibition increased as the CP leaf extract's concentration increased. However, it mistakenly fluctuates at 125 µg/mL concentration. Interestingly, the maximum inhibition towards 3CLpro was reached at the fifth tested concentration of 750 µg/mL. This could describe the maximum velocity of the 3CLpro to catalyze the proteolytic reaction that occurred at the previous concentration of 500 µg/mL. The $IC_{50}$ of CP leaf ethanolic extract was observed at 0.02 µg/mL ($R^2$ = 0.6128).

On the other hand, CP leaf extract showed 29% inhibition at the lowest concentration (125 µg/mL) against SARS-CoV-2 PLpro. The inhibition increases along with increasing the concentration up to 81.84%. However, at the fourth concentration (750 µg/mL), the inhibition started to decrease, which means that it reached its maximum capacity. The $IC_{50}$ of CP leaf extract towards PLpro was not far away from the 3CLpro, which equals to 0.06 µg/mL ($R^2$ = 0.4396).

In the TMPRSS2, the CP leaf extract shows 35% inhibition at its lowest concentration of 125 µg/mL, but it decreased at the fifth concentration (1000 µg/mL). The maximum inhibition, which equals 87%, occurred at 750 µg/mL. The $IC_{50}$ of CP leaf extract towards TMPRSS2 was calculated at 1,888 µg/mL with an $R^2$ value of 0.6774. This confirms the weak activity of such a sample towards TMPRSS2 as a target. **Fig. 2** illustrates the dose-dependent curves of CP leaf extract against SARS-CoV-2 3CLpro, PLpro and the human TMPRSS2.

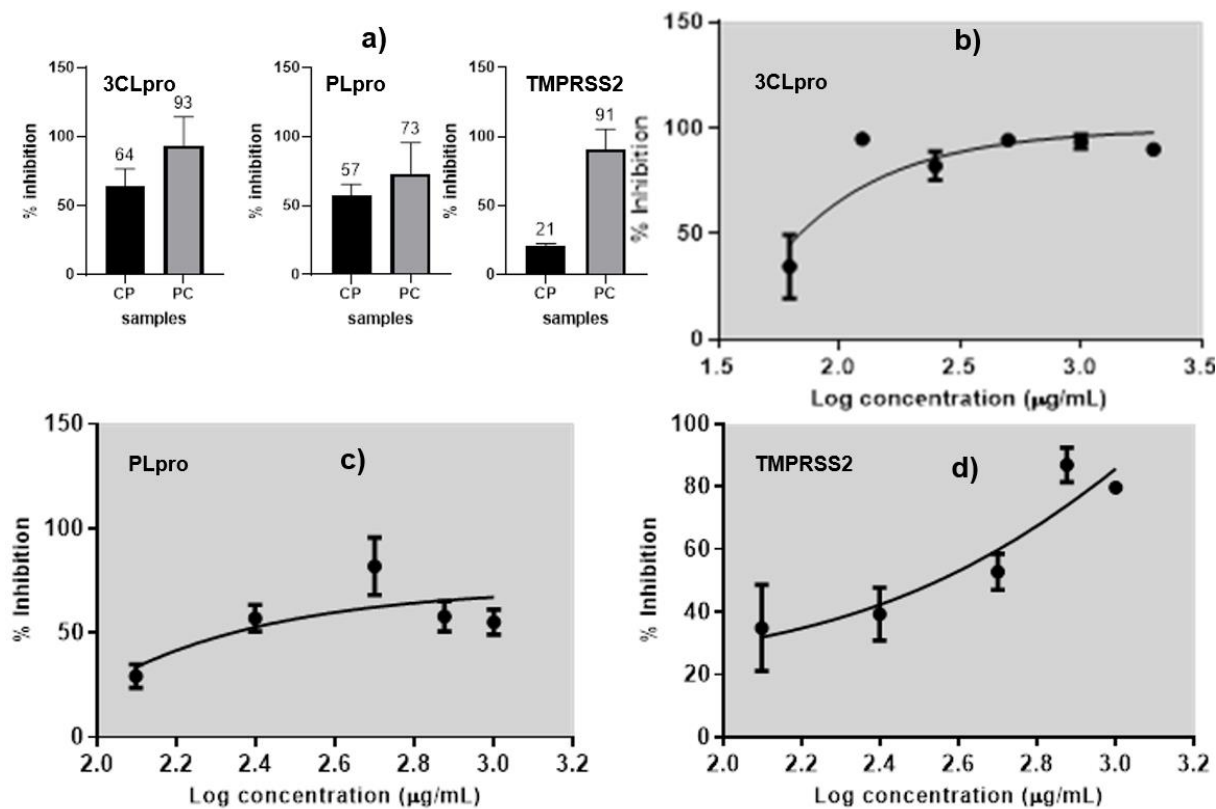

**Fig. 2** The inhibition profiles of CP leaf extract represented by **a)** histograms comparing the % inhibition of CP leaf extract (64% at 250 µg/mL) with the positive control (PC; GC376) (93% at 100 µM) against 3CLpro, CP leaf extract (57% at 250 µgmL) with the positive control (PC; GRL0617) (73% at 100 µM) against PLpro, and CP leaf extract (21% at 250 µg/mL) with the positive control (PC; camostat) (91% at 100 µM), followed by the dose-dependent curves against **b)** SARS-CoV-2 3CLpro with an $IC_{50}$ of 0.02 µg/mL ($R^2 = 0.6128$), **c)** PLpro with an $IC_{50}$ of 0.06 µg/mL ($R^2 = 0.4396$), and **d)** TMPRSS2 with an $IC_{50}$ of 1,888 µg/mL ($R^2 = 0.6774$).

The cytotoxicity study of CP leaf extract against Vero cells demonstrates an average $CC_{50}$ 1,317 µg/mL, confirming that the CP leaf extract is non-toxic. This is due to its safety index (SI), which was calculated as the ratio between $CC_{50}$ and $IC_{50}$ of the extract against 3CLpro and PLpro (65.80 and 21.90, respectively). However, the SI against the TMPRSS2 might not be considered since CP leaf extract was inactive against this enzyme. The dose-dependent curve of CP leaf extract tested in Vero cells is presented in **Fig. 3a**. Cells' morphology for the negative control and CP leaf extract treatments upon MTT assay is illustrated in **Fig. 3b** and **c**, respectively.

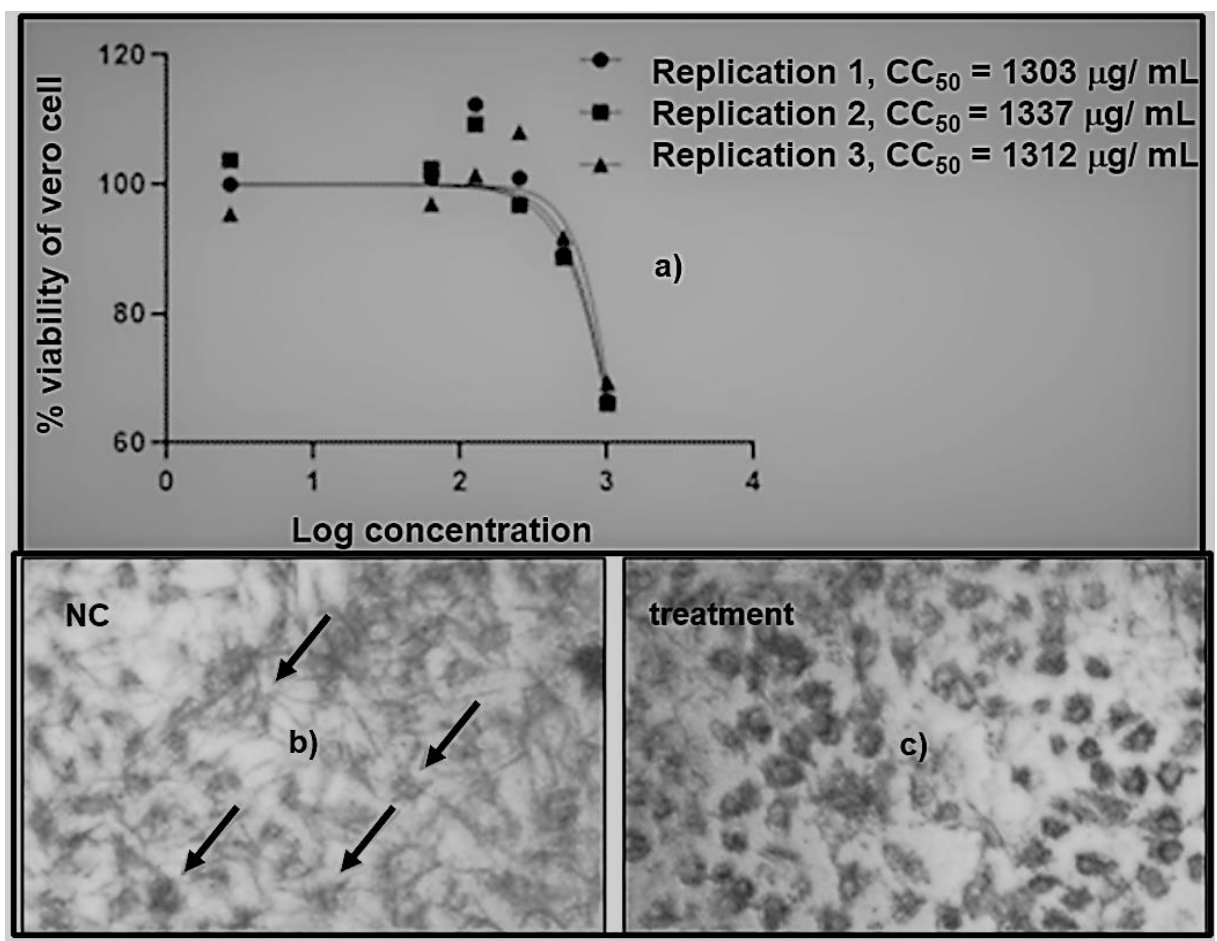

**Fig. 3** The cytotoxicity profile of CP leaf extract presenting **a)** dose-dependent curves from the triplicated experiment with an average $IC_{50}$ of 1,317 μg/mL, **b)** Vero cells morphology as in the negative control (NC) producing formazan crystals upon MTT reaction indicated by black arrows, and **c)** Vero cells morphology treated with the CP leaf extract reduced the production of formazan crystals.

The GC chromatogram (**Fig. 4a**) identified 55 peaks (**1a-55a**), which are detailed in **Table 2S**. This experiment only detects the volatile compounds, which have not been previously discussed in our published articles, i.e., flavonoids, flavonoid glycosides, cyanogenic glycosides, coumarins, quinones, cinnamic acids, phenols, and alkaloids [22, 4]. GC-MS observed two major compounds having relative areas of 32.41% and 3.82% for peaks **17a** and **21a**, respectively. Although, there are three other major peaks; **10a** (12.57%), **12a** (8.63%), and **13a** (6.72%), showing higher relative areas (in parentheses) than peak **21a**, we rather picked up peak **21a** due to its better resolution. The MS detects peaks **17a** and **21a** with their molecular weights/ion ($m/z$) ratios of 218.0 and 384.5, respectively, as illustrated in **Fig. 4b** and **c**. The $m/z$ was then matched to the NIST library having a similarity index close to phenol, 2-methyl-5-(1,2,2-trimethylcyclopentyl)-(*S*)- and β-mannofuranoside, farnesyl- for $m/z$ 218.0 and $m/z$ 384.5, respectively.

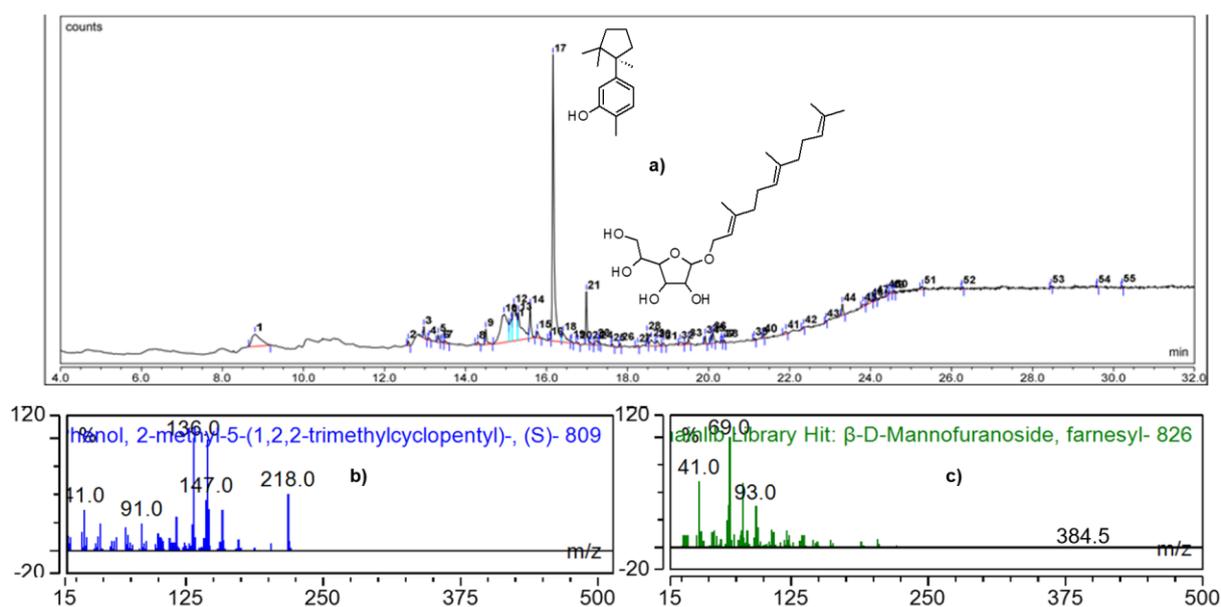

**Fig. 4** The phytochemical profiles of the CP leaf extract represented by **a)** GC chromatogram of 55 peaks associated with 55 identified compounds, and the mass spectra of **b) 17a** showing an $m/z$ of 218.0, which corresponds to phenol, 2-methyl-5-(1,2,2-trimethylcyclopentyl)-(*S*)-, whereas **c) 21a** showing an $m/z$ of 384.5, which corresponds to β-mannofuranoside, farnesyl-.

On the other hand, more general compounds, either volatile or non-volatile, were detected via LC-MS by targeting carpaine (m.p. 121°C; MW 478.70 g/mol), one of the alkaloids reported in CP leaves. The LC chromatogram (**Fig. 5a**) shows eight peaks (**1b-8b**) at $R_t$'s of 0.29, 2.81, 4.02, 6.82, 14.27, 15.27, 15.85, and 16.27 min. As expected, carpaine presents at 4.02 min (**3b**) showing an *m/z* of 479.8872 (M+H) in the positive ion mode (**Fig. 5b**). A few other compounds reported in CP are also present in the MS spectrum such as 5,7-dimethoxycoumarin at 0.29 min (**1b**; *m/z* 229.4762 (M+Na)) (**Fig. 5c**), anthraquinone at 6.82 min (**4b**; *m/z* 230.5112 (M+Na)) (**Fig. 1Sa**), and niacin at 15.85 min (**7b**; *m/z* 123.4189 (M)) (**Fig. 1Sb**).

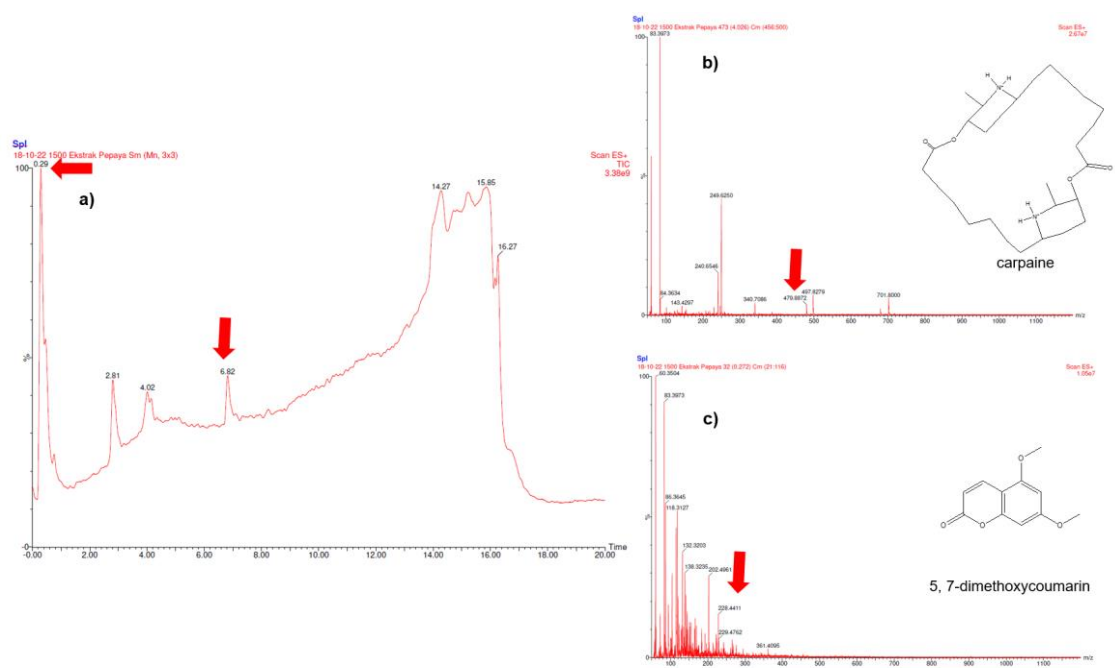

**Fig. 5** The phytochemical profiles of the CP leaf extract represented by **a)** LC chromatogram of eight peaks associated with at least two identified compounds at 0.29 min and 6.82 min (red arrows), and the mass spectra of **b) 3b** showing an *m/z* of 479.8872 (M+H) which is identified as carpaine, whereas **c) 1b** showing an *m/z* of 229.4762 (M+Na), which matches the hit 5,7-dimethoxycoumarin.

The mechanism on how the two identified compounds interact with the individual protein targets is studied using molecular docking, in which the results are tabulated in **Table 1**. In the docking result of 3CLpro, **17a** (-6.94 kcal/mol), **21a** (-7.14 kcal/mol), and **3b** (-8.45 kcal/mol) showed slightly lower $\Delta G_{bind}$ in comparison to baicalein (-6.38 kcal/mol; the co-crystallized ligand). In contrast, **17a** (-6.06 kcal/mol), **21a** (-5.91 kcal/mol), **1b** (-5.43 kcal/mol), and **3b** (-5.97 kcal/mol) showed slightly higher $\Delta G_{bind}$ compared to VIR251 (-7.00 kcal/mol; the co-crystallized ligand of PLpro). However, due to the close $\Delta G_{bind}$ range, they could be considered with the same potency as the main proteases' inhibitors. Next, the binding of **17a** (-6.53 kcal/mol), **21a** (-5.44 kcal/mol), and **1b** (-5.90 kcal/mol) with the TMPRSS2 also resulted in slightly higher $\Delta G_{bind}$ compared to nafamostat (-6.40 kcal/mol), but it is also in a close range in the binding energy.

**Table 1**   The $\Delta G_{bind}$ and interacting residues of baicalein, VIR251, nafamostat, **17a**, and **21a** with SARS-CoV-2 main proteases and the human TMPRSS2 based on the docking study.

| Ligands | 3CLpro | | PLpro | | TMPRSS2 | |
|---|---|---|---|---|---|---|
| | $\Delta G_{bind}$ (kcal/mol) | Interacting Residues | $\Delta G_{bind}$ (kcal/mol) | Interacting Residues | $\Delta G_{bind}$ (kcal/mol) | Interacting Residues |
| **Baicalein** | -6.38 | HIS41, GLY143, ASN142, CYS145 GLU166 | na | na | na | na |
| **VIR251** | na | na | -7.00 | GLY163, GLY271, ASP164, ASN109 | na | na |
| **Nafamostat** | na | na | na | na | -6.40 | GLY439, ASP440, SER441, SER436, TRP461, GLY464, ASP435, ARG470, PRO471 |
| **17a** | -6.94 | GLU166 | -6.06 | ASP164 | -6.53 | SER436, VAL473, GLY472 |
| **21a** | -7.14 | GLU166, CYS145, MET165 | -5.91 | ASP164, ARG166, PRO247, TYR264, TYR273 | -5.44 | VAL473, GLY472, ASP435, GLY464, SER436 |
| **1b** | -5.86 | GLY143, GLU166 | -5.43 | TYR273 | -5.90 | na |
| **3b** | -8.45 | CYS44 | -5.97 | TYR264 | -6.48 | GLY462, GLY464 |

na, not applied

The binding energy of the ligands in 3CLpro are mostly contributed by the H-bond interactions, in which **17a**, **21a**, **1b,** and **3b** exhibited at least one interaction that also exists in the reference ligands. **Fig. 6a** and **b** show the docking poses of **17a** and **21a** at the binding site of 3CLpro while interacting with GLU166, and GLU166 & CYS145, respectively. As seen earlier, **21a** showed more H-bond interactions due to its higher numbers of HBDs and HBAs, which contributes to its lower $\Delta G_{bind}$ compared to the other three compounds. The interaction of these ligands with GLU166 is important because this binding mode is also present in baicalein. The binding of **21a** with 3CLpro is tightened by its interaction with CYS145, one of the catalytic dyads in 3CLpro. This could explain how CP leaf extract is able to interfere with 3CLpro through the enzymatic bioassay. On the other hand, **3b** (**Fig. 6c**) shows lower energy of binding than the other three ligands as well as baicalein. However, no H-bond interaction was noticed with any important residues that contributed to this binding energy. Only one H-bond interaction was seen with CYS44 with an atomic distance of 3.0 Å. Meanwhile, **1b** (**Fig. 6d**) was able to interact with GLY143 and GLU166. Somehow, the binding energy is higher than the other ligands. Instead, the hydrophobic parts of **17a**, **21a**, **1b**, and **3b,** such as the trimethylcyclopentyl chain, farnesyl chain, long alkyl chains, and the methoxy groups could contribute to the binding energy via non-bonding interactions, such as van der Waals (vdW), carbon H-bond, pi-cation, alkyl, and pi-alkyl, as illustrated in **Fig. 2S**.

The binding of **17a** and **21a** to PLpro showed important H-bond interactions with ASP164, which is also present in VIR251 (**Fig. 6e** through **f**). Although no interactions were found with any of the catalytic triad residues (CYS111, HIS272, and ASP286) [23], further assessment of the non-bonding interactions revealed that **21a** interacts with CYS111 via a pi-alkyl interaction (**Fig. 3S**). Again, due to the higher HBD and HBA counts in **21a**, extra H-bond interactions were noticed, especially with ARG166, PRO247, TYR264, and TYR273. However, these extra binding interactions are relatively weak due to their longer atomic distances (2.1-3.3 Å), in comparison to ASP164 (1.7 Å), which weakly contributes to the $\Delta G_{bind}$ calculation. On the other hand, **3b** and **1b** (**Fig. 6g** and **h**) interacted only with TYR273 and TYR264, respectively, which are not represented by the binding of VIR251. Therefore, the binding modes of **1b** and **3b** in PLpro are not significantly specific.

The docking poses of **17a**, **21a**, and **3b** at the binding site of TMPRSS2 are illustrated in **Fig. 6i** through **k**, showing similar binding modes to that of nafamostat by interacting with SER436, VAL473, and GLY464. Another two extra binding residues were present in the case of **21a**, which include ASP435 and GLY464. However, these two binding interactions seem weak interactions and did not contribute to the $\Delta G_{bind}$ calculation of **21a**. Further evaluation showed that **1b** (**Fig. 6l**) did not interact with TMPRSS2 via any H-bond interaction, supporting the *in-vitro* results described earlier. These interesting poses show that both the hydrophilic and the hydrophobic parts of the two ligands (**21a** and **1b**) always occur in the same orientation. In addition, the non-bonding interaction of the hydrophobic parts of **17a**, **21a**, **3b**, and **1b** with TMPRSS2 was presented in **Fig. 4S**.

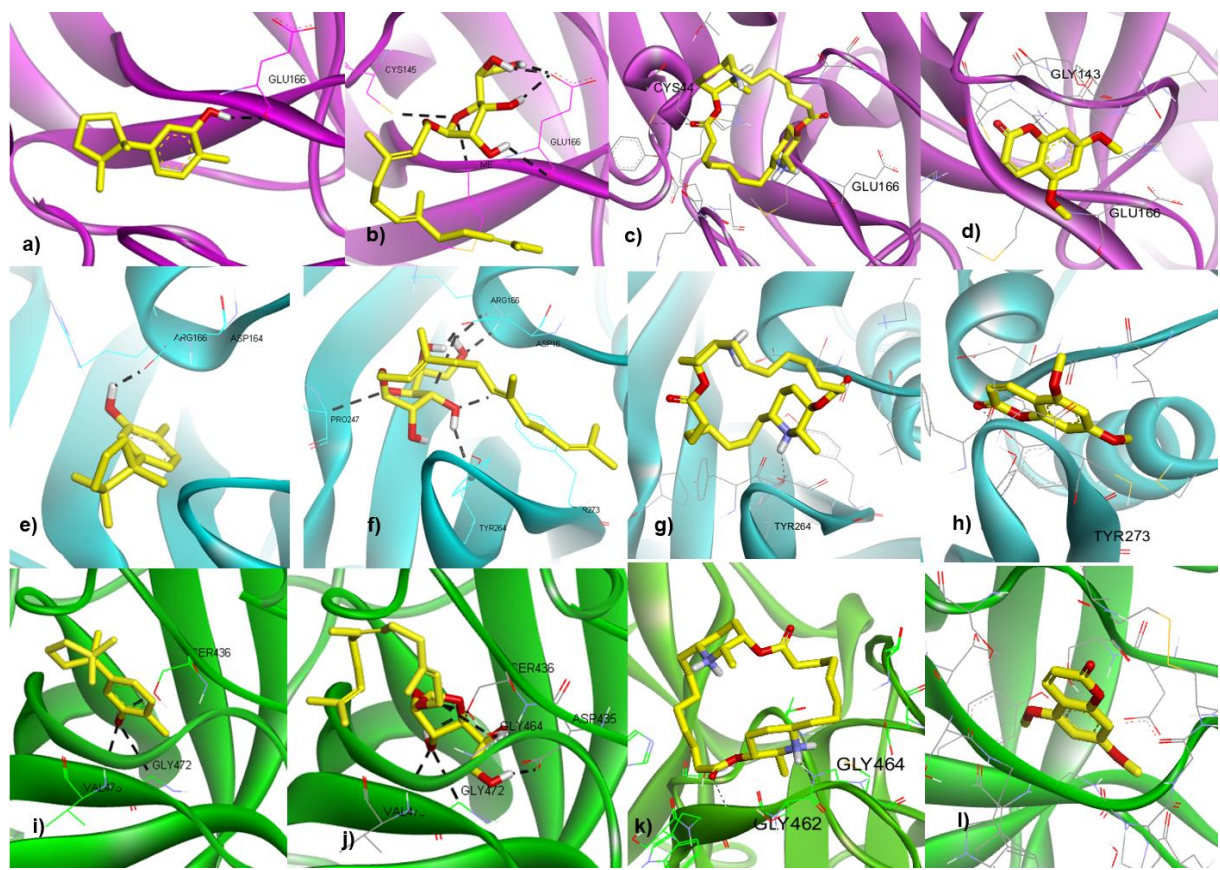

**Fig. 6** The docking poses of **17a**, **21a**, **3b**, and **1b** in the binding site of **a-d)** SARS-CoV-2 3CLpro, **e-h)** SARS-CoV-2 PLpro, and **i-l)** the human TMPRSS2, respectively.

Compounds **17a**, **21a**, **3b**, and **1b** were screened against the pharmacophore model of baicalein co-crystallized with 3CLpro. PLpro and TMPRSS2 were not studied for their pharmacophores at the moment, as they have not been validated against external libraries. The 3CLpro pharmacophore model has a pentagon shape which is composed of three hydrogen bond acceptor (HBA) features and two hydrophobic (H) features with various inter-distances as follows: H-H (6.15 Å), H-HBA (2.38 Å), HBA-HBA (2.76 Å), HBA-HNA (5.14 Å), and HBA-H (6.93 Å) [24]. Compounds **17a** and **21a** were able to fit in the pharmacophore model with fit scores of 46.24 and 46.03, respectively. Compound **17a** fits to two hydrophobic (H) features, which are represented by the methyl group and the phenyl ring along with two hydrogen bond acceptors (HBAs), which are represented by the OH group. On the other hand, compound **21a** is, unfortunately, missing the H features. However, it still fits the four HBA features through one of the OH groups. The pharmacophore features of **17a** and **21a** are illustrated in **Fig. 7**.

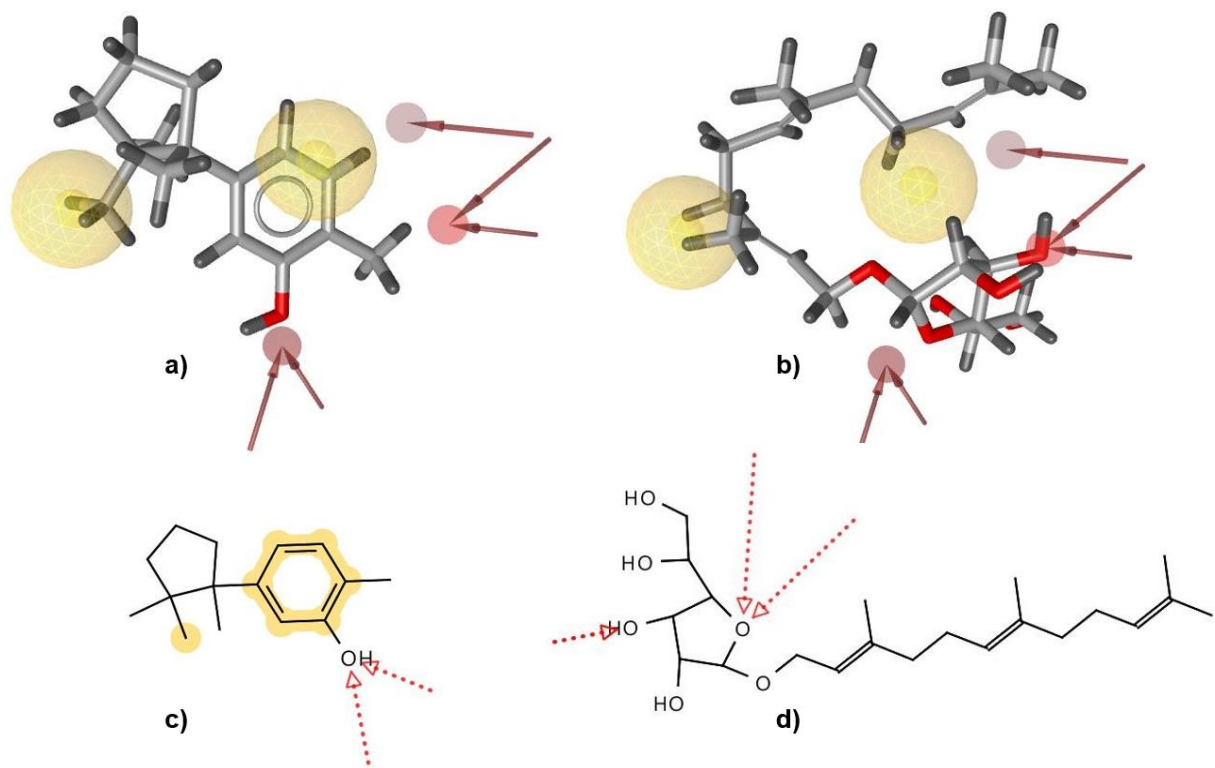

**Fig. 7** The pharmacophoric profiles of the two identified compounds (**17a** and **21a**) as SARS-CoV-2 3CLpro inhibitors presenting **a) 17a**, and **b) 21a** in 3D structures along with **c) 17a** and **d) 21a** in 2D. The yellow spheres represent the H feature, whereas the red spheres and the red arrows (and the dashed arrows) represent the HBA feature.

Compound **17a** is a phenol extended by 2-methyl-5-(1,2,2-trimethylcyclopentyl)-(*S*)-. The phenol group drives the HBD to GLU166, which is important for interrupting the 3CLpro activity. This also occurs in baicalein, although such a binding mode was driven by the carbonyl O of the cyclohexanone ring. Other phenol compounds were reported as SARS-CoV-2 3CLpro inhibitors, such as quercetin, ellagic acid, curcumin, epigallocathecin, and resveratrol, showing

IC$_{50}$'s ranging between 11.8 and 23.4 µM [25]. Instead, non-phenolic glycosides, such as β-mannofuranoside, farnesyl- (**21a**), could act as an inhibitor of the corresponding enzyme. This compound is more remarkable, due to its lower ΔG$_{bind}$, in comparison to **17a**. Interestingly, this compound was able to interact with CYS141, one of the catalytic dyads in the active site of 3CLpro. In addition, the flavonoid glycosides of baicalein and pectolinarin were reported as 3CLpro inhibitor accordingly [26]. Further compounds, i.e., **3b** and **1b,** do not fit to the pharmacophore model in its given parameters. These findings consistently agree with the docking result, in which **3b** did not interact with any important residue of the 3CLpro. Although the binding energy is slightly lower, vdW and other non-bonding interactions are predominant. Meanwhile, **1b** showed the highest binding energy among all ligands. Therefore, it is unfavorable to be proposed as a lead inhibitor against SARS-CoV-2 3CLpro.

Not so far behind, CP leaf extract also showed a significant inhibition toward SARS-CoV-2 PLpro with an IC$_{50}$ of 0.06 µg/mL. A typical protease inhibition is performed by a phenol compound, such as **17a**. This compound interacts via H-bond interactions with ASP164, which also presents in the co-crystallized ligand (VIR251), although this binding is driven by a different HBA; the acetamide N. Moreover, the antihelmintic pyrantel pamoate also demonstrated its activity towards SARS-CoV-2 PLpro with an IC$_{50}$ of >20 µM through its phenol groups [27]. In VIR251, the presence of a tyrosine side chain could support the essential role of phenolic compounds in disrupting PLpro activity. The PLpro inhibitor is identically described by a covalent bond between the amide group of the ligand with CYS111. Unfortunately, this irreversible binding is generally unfavorable due to the toxic potential [28]. The absence of a covalent bond between **17a** and **21a**, in this case, could offer an inhibitor of PLpro with less toxic effects.

As seen earlier, CP leaf extract demonstrated a weak inhibitory activity toward human TMPRSS by showing an IC$_{50}$ of 1888 µg/ mL only. This *in vitro* result seems correlating with the docking result demonstrating ΔG$_{bind}$ of **17a** and **21a** toward human TMPRSS2 are -6.53 kcal/ mol and -5.44 kcal/mol, respectively, slightly higher than their ΔG$_{bind}$ in 3CLpro. Nafamostat and camostat are two strong serine protease inhibitors due to their covalent bond through the SER441 phenylguanidino acylation in human TMPRSS2 [21]. This residue is one of the catalytic triads, next to HIS296 and ASP345 in human TMPRSS2. Although **17a** and **21a** are able to interact with important residues of the TMPRSS2, however, their lacking acyl-ester group could probably the reason to weakly active towards this protein. One recent study reported 3-amidinophenylalanine-derived inhibitors, MI-432 and MI-1900, as human TMPRSS2 inhibitors due to the presence of amidinophenylalanine, which mimics the phenylguanidine tandems with acyl amides in their molecules [29-31].

The *in-vivo* acute/chronic toxicological studies confirm that no toxic effect performed by the CP leaf extract, which support our *in-vitro* cytotoxicity study [32, 33]. The pharmacophore mapping of the two identified compounds, demonstrating that **17a** is more fit to the pharmacophore model of SARS-CoV-2 3CLpro inhibitor than **21a**. The higher fit score and the pharmacophoric features of **17a** than **21a** could suggest that the major compound responsible for the 3CLpro inhibition during the in-*vitro* assay is **17a**. The pharmacophore of **17a** also deals with its docking pose, in which the OH group serves as HBD as well as HBA toward GLU166 along with vdW interactions between the methyl group and TYR52, ASP187, and ARG188. At the same time, the phenyl ring interacts with MET165 via a pi-alkyl interaction, another type of the hydrophobic interactions, instead of vdW. Further studies should include the pharmacophore profiles of **17a** and **21a** against the PLpro since CP leaf extract is also active against this enzyme.

## 3. CONCLUSIONS

This study reports the *in-vitro* and *in-silico* activity of *Carica papaya* (CP) ethanolic leaf extract towards the two proteases of SARS-CoV-2, 3CLpro and PLpro, and the human TMPRSS2 which facilitates for the virus attachment to host's cells. First, the extract was *in-vitro* assayed against all three protein targets, where results showed a more significant activity of the extract towards the two proteases represented by the $IC_{50}$ values of 0.02 and 0.06 µg/mL for 3CLpro and PLpro, respectively, in comparison to TMPRSS2 ($IC_{50}$ = 1,888 µg/mL). Moreover, the extract was found to be non-toxic towards Vero cells with a $CC_{50}$ of 1,317 µg/mL. This has led us to investigate further the phytochemical components of the extract via GC-MS and LC-MS. Through GC-MS, the most pronounced peaks were found to belong to phenol, 2-methyl-5-(1,2,2-trimethylcyclopentyl)-(S)- (**17a**) and β-mannofuranoside, farnesyl- (**21a**), while LC-MS confirmed the presence of carpaine, 5,7-dimethoxycoumarin, anthraquinone, and niacin.

Findings from the molecular docking confirmed the *in-vitro* results, where three out of the four selected compounds possessed $\Delta G_{bind}$ lower than that of the reference compound in 3CLpro. Additionally, and in the case of 3CLpro, **17a** and **21a** were found to interact with two essential amino acid residues which are responsible for the protease catalytic activity. On the other hand, none of the docked compounds displayed any significant H-bond interactions with amino acids of the PLpro or TMPRSS2 which might contribute to their inhibitory activity towards these enzymes. The weak inhibitory activity of **17a** and **21a** towards TMPRSS2 was contributed to their lack of an acyl-ester group which is essential for the inhibition of the enzyme. Finally, the same four compounds were screened against a baicalein co-crystallized with 3CLpro pharmacophore model. Compounds **17a** and **21a** exhibited fit scores of 46.24 and 46.03, respectively. Compound **17a** was found to fit to two hydrophobic (H) and two hydrogen bond acceptors (HBAs), whereas **21a** was missing the H features and only fitted to four HBA features. Still, further investigations are required for these two compounds towards the 3CLpro through a time-evolution dynamic study, and against a pharmacophoric model of PLpro.

## 4. MATERIAL AND METHODS

### 4.1 Chemicals

The CP leaf extract was supplied by PT Industri Jamu and Farmasi Sido Muncul Tbk., Indonesia. The 3CLpro, PLpro and TMPRSS2 fluorogenic assay kits were purchased from BPS Bioscience, USA. The 3CLpro kit contains recombinant 3-chymotrypsin-like pro, DABCYL-KTSAVLQSGFRKME-EDANS substrate, 3CLpro buffer, (2$R$)-1-hydroxy-2-[[(2$S$)-4-methyl-2-(phenylmethoxycarbonylamino)pentanoyl]amino]-3-(2-oxopyrrolidin-3-yl)propane-1-sulfonate sodium (GC376) as the positive control and 1,2-1,4-Dithiothreitol (DTT). The PLpro assay kit contains recombinant papain-like protease, PLpro ubiquitinated substrate (RLRGG-AMC), PLpro assay buffer, 5-amino-2-methyl-$N$-[(1$R$)-1-naphthalen-1-ylethyl]benzamide (GRL0617) as the positive control, and DTT. The TMPRSS2 assay kit consists of TMPRSS2, TMPRSS2 fluorogenic substrate, 1X TMPRSS2 assay buffer, and camostat as the positive control. Vero cells were obtained from Parasitology Laboratory, Medical Faculty, Gadjah Mada University, Indonesia, and cultured in Rosswell Park Memorial Institute 1640 (RPMI 1640) medium containing 10% $v/v$ fetal bovine serum (FBS) and 1% penicillin-streptomycin (Life Technologies, USA) at 37°C in 5% $CO_2$ humidified incubator. Sodium bicarbonate ($NaHCO_3$), hydrochloric acid (HCl), sodium hydroxide (NaOH), water for injection, trypan blue, phosphate buffer saline (PBS), trypsin-ethylenediaminetetraacetic acid, doxorubicin, sodium dodecyl sulfonate (SDS), and 3-(4,5-dimethylthiazol-zyl)-2,5-diphenyl tetrazolium bromide (MTT) was obtained from Sigma, USA.

### 4.2 Computer hardware and software

An HP laptop with AMD Ryzen 5 4500U, Radeon graphics 2.38 GHz, 8 Gb RAM, and running Windows 10 and Ubuntu 2014 operating systems. The software used include Biovia Discovery Studio 2021 Visualizer (3ds.com), MGLTools 1.5.6 (ccsb.scripps.edu), Ligandscout 4.4.7 (inteligand.com), and GraphPad Prism 9 (graphpad.com).

**4.3 The 3CLpro FRET-based assay**

A solution of 0.5 M DTT was diluted 500 times into 3CLpro assay buffer to obtain a 1 mM concentration and stored at -20°C. The 3CLpro was thawed on ice, briefly spun, and diluted in the assay buffer containing 1 mM DTT at 0.5 ng/mL (15 ng per reaction). Thirty (30) µL diluted 3CLpro was added to the wells and designated as "positive control", "inhibitor control", and "test sample", followed by adding 30 µL of assay buffer to the "blank" wells. On the other hand, 50 μg GC376 was dissolved in 200 μL assay buffer to obtain 500 μM. This solution was then added to the wells labelled as "inhibitor control" while storing the aliquot and the remaining solution at -80°C. The "test sample" was prepared by dissolving the CP leaf extract in a series of concentrations to obtain 62.5, 125, 250, 500, 750, and 1000 µg/mL as the final concentrations with 1% *v/v* DMSO-buffer as the solvent, followed by pipetting 10 µL of the prepared samples into the wells. Ten (10) µL of the DMSO was added to the wells and assigned as a "diluent inhibitor". The mixture was pre-incubated for 10 min at room temperature while slowly shaking. The substrate (10 μL; 200 μM) was added to wells to obtain 40 μM as the final concentration. The mixture was then incubated at room temperature for 4 hours. The assay was run in triplicate, and its fluorescence was measured using Synergy HTX multimode reader UV optical kit at λ 480/535 nm. The "positive control" is named for the reaction between enzyme and substrate without inhibitor [34-36].

**4.4 The PLpro FRET-based assay**

Thirty (30 μL) of diluted PLpro (0.3-0.5 ng/μL in the assay buffer containing 0.5 μM DTT) was added to the wells and designated as "positive control", "inhibitor control", and "test sample", followed by adding 30 µL of the assay buffer to the "blank" wells. The tested sample of CP leaf extract was prepared in a series of concentrations (125, 250, 500, 750, and 1000 µg/mL) in 1% DMSO and stored at -20°C for further use. The sample (10 µL) was then dispensed into the wells, along with 30 µL of PLpro (500 µM), and incubated for 30 min at 37ºC. Ten (10) µL of the "inhibitor control" (GRL0617 500 µM) was added to the wells to obtain 100 µM as the final concentration while storing the aliquot and the remaining solution at -80°C. The reaction was initiated by adding 10 µL of 250 µM substrate to obtain 25 µM as the final concentration and incubated for 60 min at 37ºC, followed by measuring its fluorescence using Synergy HTX multimode reader UV optical kit at λ 360/460 nm. The experiment was run in duplicate [37-39].

### 4.5 The TMPRSS2 FRET-based assay

The activity of TMPRSS2 was measured using the fluorogenic substrate (78047) in a black 96-microwell plate. TMPRSS2 was dissolved in 1x TMPRSS2 assay buffer up to 5 ng/mL or 150 ng/reaction while conditioned at a cold temperature. The "inhibitor control", camostat, was reconstituted in water for injection up to 50 mM. This solution was then 100x diluted using assay buffer to obtain 50 µM. Ten (10) µL of CP leaf extract's prepared concentrations (125, 250, 500, 750, and 1000 µg/mL in 5% DMSO) were then added into the wells. The plates were incubated for 30 minutes at room temperature, and 10 µL substrate (50 µM) was then added to obtain a final concentration of 10 µM. The total volume of the reaction was 50 µL per well, and the plates were re-incubated for an extra 10 min. The assay was run in triplicate, and the fluorescence was measured using Synergy HTX multimode reader UV optical kit at λ 383/455 nm [40].

### 4.6 Cytotoxicity study

The series of CP leaf extract's concentrations (62.5, 125, 250, 500, 1000 µg/mL) along with the positive control (doxorubicin 12.5, 25, 50, 100, 200 µg/mL) were prepared. A standard method of cytotoxicity assay using MTT reagent was applied [41, 42]. The absorbance was read using ELISA reader (TECAN Infinite 20) at 595 nm. The cell morphology was captured using Olympus DSX1000 camera under an inverted microscope (Magnus).

### 4.7 Characterization via GC-MS and LC-MS

Sample (±100 mg) was dissolved in 1.5 mL 96% ethanol and centrifuged at 9500 rpm for 3 min. A volume of 0.5 µL of the solution was injected into the GC (Thermo Scientific Trace 1310 Gas Chromatograph) with HP-5MS UI column. The GC is coupled to a Mass Spectrometer (Thermo Scientific ISQLT Single Quadrupole Mass Spectrometer). Helium gas was used a carrier gas, and the GC oven was initially held at 100°C for 5 min, and then elevated 5°C per min to reach 230°C. Peaks in the chromatograms produced by this analysis were identified by a combination of references to their mass spectra and the NIST08 mass spectral database.

On the other hand, to run the LC-MS analysis, a sufficient amount of the sample was dissolved in ethanol and filtered through millex 0.22 µm. Five (5) µL of the sample was injected into HPLC with conditions as follows: column UPLC BEH C18, 1.7 µm, 2.1×50 mm; mobile phase water formic acid 0.1% (A) and acetonitrile formic acid 0.1% (B) with a gradient system as tabulated in **Table 2**. The MS was conditioned at a mass range of 50-1200, cone voltage of 30 V, capillary voltage of 3.0 kV, source temperature of 500°C, and a positive polarity mode.

**Table 2.** The gradient system of HPLC in identifying the general compounds of CP extract.

| Time | Flow | %A | %B |
|------|------|-----|-----|
| 0    | 0.4  | 95  | 5   |
| 15   | 0.4  | 0   | 100 |
| 15.1 | 0.4  | 95  | 5   |
| 20   | 0.4  | 95  | 5   |

### 4.8 Molecular docking

The protein was retrieved from the Protein Data Bank (rcsb.com) with the PDB codes as follows: 6M2N for SARS-CoV-2 3CLpro [19], 6WX4 for SARS-CoV-3 PLpro [20], and 7MEQ for the human TMPRSS [21]. The docking coordinates were centered on the ligand as follows: 3CLpro (-33.362, -65.436, and 41.436, as $x,y,z$), PLpro (9.508, -27.455, and -37.505, as $x,y,z$), and TMPRSS2 (-9.831, -6.278, and 20.329, as $x,y,z$). The grid box volume was set at 40×40×40 points with a 0.375 grid spacing. The docking was run for 100 Lamarckian Genetic Algorithm runs with a 150 population size and 2,500,000 as the number of evaluations packed in AutoDock4 [43]. The validation is defined when the redocking shows root mean square deviation (RMSD) from their initial pose with not greater than 2 Å (**Fig. 4S**) [44]. The ligands (**17** and **21**) chemical structures were downloaded from PubChem (pubchem.com) and converted into 3D structures using Biovia Discovery Studio.

### 4.9 Pharmacophore mapping

The pharmacophore model of baicalein bound to SARS-CoV-2 3CLpro (PDB 6M2N), generated and validated in our previous published study [45] was used as the query. The ligands were then screened into the individual pharmacophore using the screening tool with the default parameters [45] except for the maximum number of omitted features, which was set to three features.

**Supporting Information**

The triplicate inhibition (%) of CP leaf extract against SARS-CoV-2 3CLpro, SARS-CoV-2 PLpro, and human TMPRSS2 are available in **Table 1S**. The 55 compounds identified using GC-MS from CP leaf extract is available in **Table 2S**. The mass spectrum of **4b** and **7b** are illustrated in **Fig. 1S**. The non-bonding interaction of a) **17a** and b) **21a** docking poses in the binding pocket of SARS-CoV-2 3CLpro-PLpro, and human TMPRSS2 are available in **Fig. 2S-4S**. The internal validation (re-docking) poses of baicalein, VIR251, and nafamostat in the binding pocket of SARS-CoV-2 3CLpro-PLpro, and human TMPRSS2 are available in **Fig. 5S**.

**Acknowledgments**

This project was financially supported by the General Director of Higher Education, Indonesian Ministry of Education, Culture, Research and Technology under the Free of Learning-Free Campus Grant (Hibah Merdeka Belajar Kampus Merdeka) 2021, No. 50/E1/KM.05.03/2021. We acknowledge Dr. apt. Rifky Febriansah, M.Sc. for providing the cell culture laboratory facility during cytotoxicity assay in Faculty of Medicine, Universitas Muhammadiyah Yogyakarta, Indonesia and GraphPad Software Company for facilitating a free trial of GraphPad Prism 9.

**Conflicts of Interest**

The authors declare no conflicts of interest.